\documentclass[a4paper,11pt]{article}
\usepackage{pos}
\usepackage{lineno}
\newcommand{\pp}{\ensuremath{\mathrm {p\kern-0.05em p}}}
\newcommand{\PbPb}{\ensuremath{\mbox{Pb--Pb}}}
\newcommand{\GeVc}{\ensuremath{\mathrm{GeV}\kern-0.05em/\kern-0.02em c}}

\newcommand{\pT}{\ensuremath{p_{\mathrm{T}}}}
\newcommand{\kT}{\ensuremath{k_{\mathrm{T}}}}

\newcommand{\pTchjet}{\ensuremath{p_{\mathrm{T}}^{\mathrm{ch\; jet}}}}
\newcommand{\tg}{\ensuremath{\theta_{\mathrm{g}}}}

\newcommand{\zg}{\ensuremath{z_{\mathrm{g}}}}
\newcommand{\zcut}{\ensuremath{z_{\mathrm{cut}}}}

\newcommand{\pTsubjet}{\ensuremath{p_{\mathrm{T}}^{\mathrm{ch\; subjet}}}}
\newcommand{\zr}{\ensuremath{z_r}}

\title{Jet substructure measurements in heavy-ion collisions with ALICE}
\ShortTitle{Jet substructure measurements with ALICE}

\author{James Mulligan$^{a,b}$ for the ALICE Collaboration}
\emailAdd{james.mulligan@berkeley.edu}
\affiliation[a]{Nuclear Science Division, Lawrence Berkeley National Laboratory, Berkeley, California 94720, USA}
\affiliation[b]{Physics Department, University of California, Berkeley, CA 94720, USA}

\abstract{Jet substructure, defined by observables constructed from the distribution of constituents within a jet, provides the versatility to tailor observables to specific regions of QCD radiation phase space. This flexibility provides exciting new opportunities to study
jet quenching in heavy-ion collisions and to
ultimately help reveal the nature of the quark-gluon plasma.
The ALICE detector is particularly well-suited to jet substructure measurements
in heavy-ion collisions due to its high-precision tracking system.
In these proceedings, we report several new jet substructure measurements in Pb--Pb collisions at $\sqrt{s_{\mathrm{NN}}}=5.02$ TeV with ALICE.
These include the first fully corrected measurements of the groomed jet momentum splitting fraction, $z_{\rm{g}}$, and the groomed jet radius, $\theta_{\rm{g}} \equiv R_{\rm{g}}/R$, 
as well as $N$-subjettiness and the fragmentation distribution of reclustered sub-jets.
These measurements are compared to theoretical calculations and 
provide new constraints on the physics underlying jet quenching.
}

\FullConference{%
  *** The European Physical Society Conference on High Energy Physics (EPS-HEP2021), ***\\
  *** 26-30 July 2021 ***\\
  *** Online conference, jointly organized by Universität Hamburg and the research center DESY ***
}

\begin{document}
\maketitle

\section{Introduction}

At the Large Hadron Collider (LHC) and Relativistic Heavy Ion Collider (RHIC), 
high-energy nuclear collisions produce hot, dense droplets of a phase of matter known as the quark-gluon plasma (QGP),
consisting of deconfined quarks and gluons \cite{TheBigPicture}. 
While the interactions between the quarks and gluons in the QGP are too weak to bind them into nucleons, they remain sufficiently strong to form a strongly coupled liquid.
The microscopic structure of this liquid remains unknown, and it is not clear whether
quasiparticle degrees of freedom emerge at some scale. 
Understanding the fundamental degrees of freedom of this strongly coupled system
and how they arise from a quantum field theory is one of the major outstanding questions in quantum chromodynamics (QCD).

In order to investigate the microscopic structure of the QGP, 
one needs to probe it at a variety of length scales. 
Jets are well suited for this task, since not only can their transverse momentum span a wide range
of values but also the internal pattern of particles within jets,
known as jet substructure, enables the design of observables to target
specific regions of QCD phase space \cite{Larkoski_2020}.
The study of jet modification in heavy-ion collisions compared to proton-proton collisions,
known as jet quenching, has established that jet-medium interactions 
result in a suppression of jet yields and an excess of soft, wide angle radiation
\cite{ReviewXinNian, ReviewYacine, ReviewMajumder}.
There remain many unknowns in jet quenching theory, however, 
including the roles of color coherence, medium response, and the space-time picture of the parton shower.
Measurements of suitably chosen jet substructure observables aim to test these aspects of
jet quenching models, and ultimately provide a path to determine the microscopic nature of the QGP
using global fits (see e.g. Ref. \cite{JETSCAPE:2021ehl}).

In these proceedings, we highlight a selection of recent jet substructure results 
from the ALICE experiment \cite{aliceDetector}. 
While ALICE has performed a variety of measurements in proton-proton collisions,
which explore the transition from the perturbative to the nonperturbative regimes
(see e.g. Ref. \cite{ALICE:2021njq}), here we focus on measurements in heavy-ion collisions.
We emphasize observables that are directly comparable to theoretical models, meaning that
they are (i) analytically calculable in pQCD, and (ii) corrected for detector effects and underlying event fluctuations. 
All results presented utilize jets reconstructed from charged particles at midrapidity using the
anti-\kT{} algorithm \cite{antikt}.

\section{Groomed jet observables: \zg{} and \tg{}}

\begin{figure}[!t]
\centering{}
\includegraphics[scale=0.37]{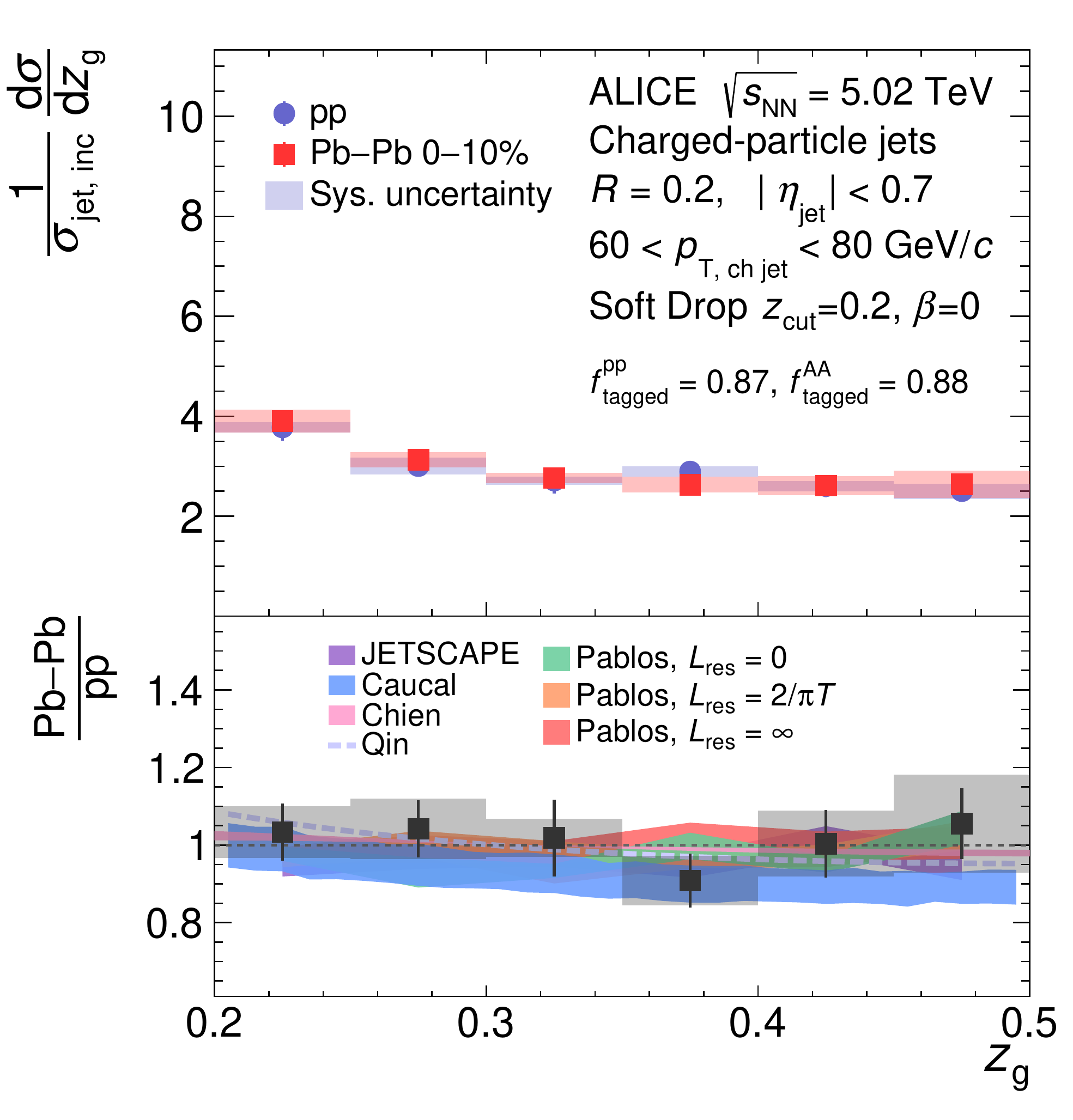}
\includegraphics[scale=0.37]{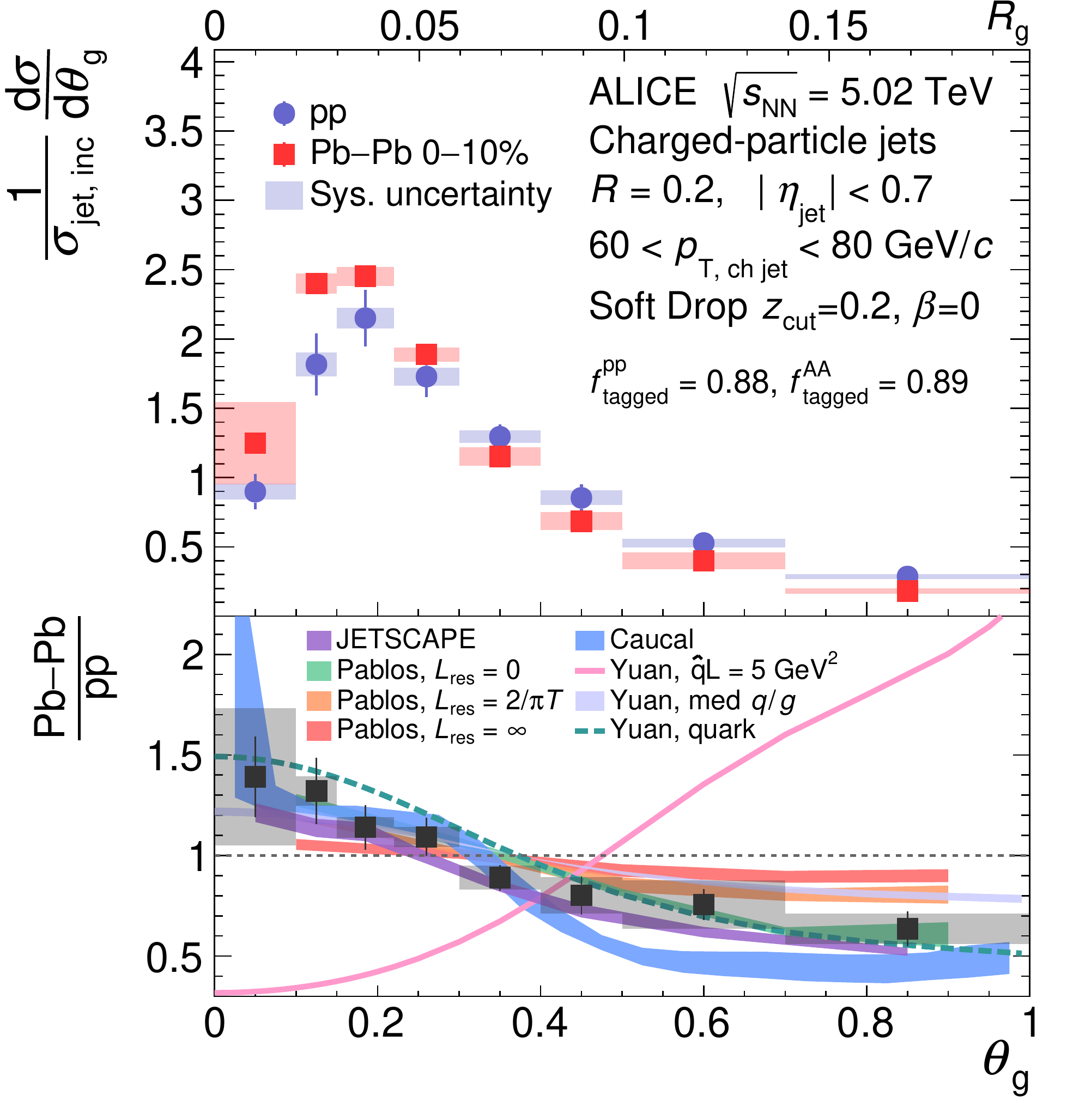}
\caption{Left: Measurements of \zg{} (left) and \tg{} (right) in \PbPb{} compared to \pp{} collisions, along with comparisons to several theoretical models \cite{ALICE:2021obz}.}
\label{fig:sd}
\end{figure}

Jet grooming techniques
have been applied to heavy-ion collisions
to explore whether jet quenching 
modifies the hard substructure of jets
\cite{PhysRevLett.119.112301, Mehtar-Tani2017, Chang:2019nrx, Elayavalli2017, Caucal:2019uvr, Ringer_2020, Casalderrey-Solana:2019ubu, Andrews_2020, PhysRevLett.120.142302, Acharya:2019djg, Sirunyan2018}.
The Soft Drop grooming algorithm identifies a single splitting 
within a jet from a grooming condition $z > \zcut \theta^\beta$, where 
$z$ is the fraction of transverse momentum carried by the sub-leading prong,
$\theta$ is the relative angular distance between the leading and sub-leading prong,
and $\beta$ and \zcut\ are tunable parameters \cite{Larkoski:2014wba, Dasgupta:2013ihk, Larkoski:2015lea}.
The groomed splitting is then characterized by two kinematic observables: 
the groomed momentum fraction, \zg{} \cite{Cal:2021fla}, and the (scaled) groomed jet radius, \tg{} \cite{Kang:2019prh}. 
By using strong grooming conditions \cite{Mulligan:2020tim}, ALICE 
measured the \zg{} and \tg{} distributions in heavy-ion collisions. 
Figure \ref{fig:sd} (left) shows that the \zg{} distribution is not significantly modified
in heavy-ion collisions,
suggesting that medium-induced radiations are not sufficiently hard to pass the grooming condition.
On the other hand, Fig. \ref{fig:sd} (right) shows a narrowing of the \tg{} distributions 
in \PbPb{} collisions relative to  \pp{} collisions \cite{ALICE:2021obz}.
These measurements are compared to a variety of jet quenching models \cite{Putschke:2019yrg, LBT, Majumder_2013, Caucal:2019uvr, Caucal_2018, PhysRevLett.119.112301, Chang:2019nrx, HybridModel, HybridModelResolution, Casalderrey-Solana:2019ubu, Ringer_2020},
most of which capture the qualitative narrowing effect observed. 
This behavior is consistent with models implementing an incoherent interaction of the 
jet shower constituents with the medium, but is also consistent with
medium-modified quark/gluon fractions with fully coherent energy loss --
presenting the opportunity for future measurements to disentangle them definitively.

\section{Sub-jet fragmentation}

In heavy-ion collisions, measurements of reclustered sub-jets have been proposed as
sensitive probes of jet quenching \cite{Kang:2017mda, Neill:2021std, Apolinario:2017qay}.
We first inclusively reconstruct jets with the anti-\kT{} jet algorithm and jet radius $R$,
and then recluster the jet constituents with the anti-\kT{} jet algorithm and sub-jet radius $r<R$.
We consider the fraction of transverse momentum carried by the
sub-jet compared to the initial jet:
$\zr = \pTsubjet / \pTchjet$.
Figure \ref{fig:zr} (left) shows the distribution of leading sub-jets with $r=0.1$ for $R=0.4$ jets 
in both \pp{} and \PbPb{} collisions. 
The distributions are compared to theoretical predictions \cite{Kang:2017mda,Qiu:2019sfj,Putschke:2019yrg, LBT, Majumder_2013} that accurately reproduce a mild rising trend of the ratio with \zr{}, which can be attributed to jet collimation. The ratio then falls as $\zr \rightarrow 1$, which may be due to the large
quark/gluon fraction at $\zr \rightarrow 1$.
These measurements offer an opportunity to probe higher $z$ than hadron fragmentation measurements,
and are an important ingredient for future tests of the universality of in-medium jet fragmentation functions.

\section{N-subjettiness}

\begin{figure}[!t]
\centering{}
\includegraphics[scale=0.37]{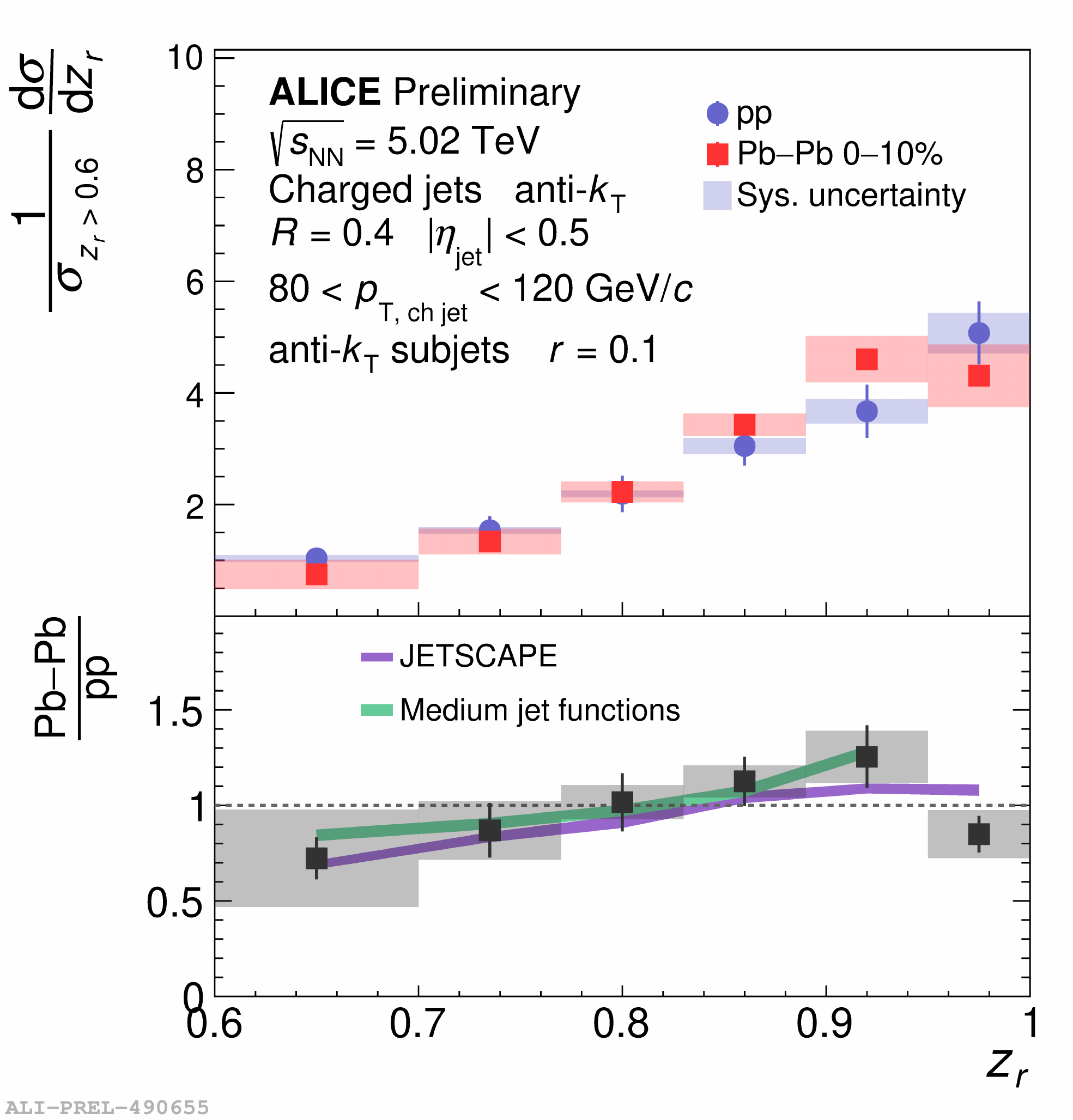}
\includegraphics[scale=0.34]{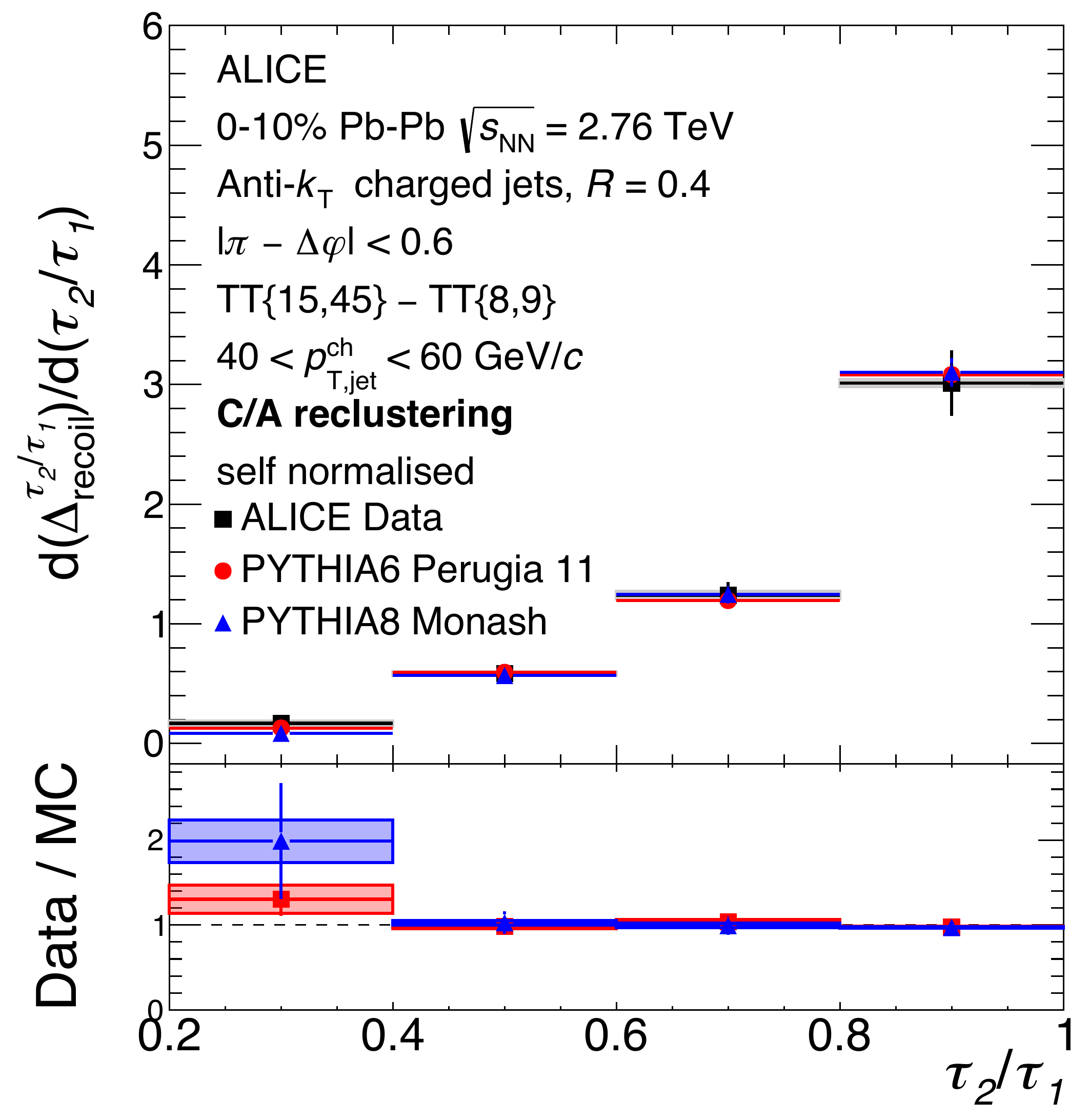}
\caption{Left: ALICE measurements of leading sub-jet fragmentation in \pp{} and \PbPb{} collisions, 
compared to theoretical predictions \cite{Kang:2017mda,Qiu:2019sfj,Putschke:2019yrg, LBT, Majumder_2013}.
Right: Measurements of the $\tau_2/\tau_1$ N-subjettiness distribution in \PbPb{} collisions  \cite{ALICE:2021vrw}
compared to PYTHIA \cite{pythia}.}
\label{fig:zr}
\end{figure}

Semi-inclusive hadron-jet correlations are well-suited to 
statistical background subtraction procedures in heavy-ion collisions, which
allows jet measurements at low \pT{} and large $R$ \cite{hjetPbPb, hjetAuAu}.
Recently, ALICE measured the N-subjettiness \cite{Stewart:2010tn,Thaler:2010tr} of jets
recoiling from a high-\pT{} hadron \cite{ALICE:2021vrw}.
Figure \ref{fig:zr} (right) shows the distribution of per-trigger semi-inclusive
yields of the $\tau_2/\tau_1$ ratio in \PbPb{} collisions compared to PYTHIA \cite{pythia}.
There is no significant modification observed in the pronginess of jets in heavy-ion collisions.
This suggests that medium-induced emissions are not sufficiently hard to produce a distinct 
secondary prong, in line with the lack of modification of the observed \zg{} distributions \cite{ALICE:2021obz}.

\section{Conclusion}

We have presented several new ALICE measurements of jet substructure in
heavy-ion collisions, which are producing an emerging picture of jet quenching phenomenology:
hard splittings are not strongly modified, as evidenced by $\zg$ and $\tau_N$, 
but there is a strong collimation or filtering effect of wide jets, as evidenced by \tg. 
The medium-induced soft splitting responsible for this filtering may be exposed in
regions dominated by quark jets, as suggested by high-\zr{} sub-jet fragmentation.
Together, these observables present opportunities for future high statistics and/or multi-differential measurements in LHC Run 3 to achieve increasingly precise constraints on jet quenching models,
and offer prospects to constrain physical properties of the QGP using global analyses.

{
\bibliographystyle{JHEP}
\bibliography{main.bib}
}

\end{document}